\documentclass[acmsmall, nonacm]{acmart}

\AtBeginDocument{%
  \providecommand\BibTeX{{%
    \normalfont B\kern-0.5em{\scshape i\kern-0.25em b}\kern-0.8em\TeX}}}

\setcopyright{none}

\usepackage{amsmath,amsfonts}
\usepackage{graphicx}
\usepackage{textcomp}
\usepackage{multirow}
\usepackage{color}
\usepackage[ruled, vlined]{algorithm2e}
\usepackage{epstopdf}
\usepackage{comment}
\usepackage{multirow}
\usepackage{url}

\newcommand{\revision}[1]{\textcolor{black}{ #1}}

\newcommand{\our}{\textit{Ridergo\,}}

\begin{document}

\title{Impact of Driving Behavior on Commuter's Comfort during Cab Rides: Towards a New Perspective of Driver Rating}

\author{Rohit Verma}
\email{rv355@cam.ac.uk}
\affiliation{%
 \institution{University of Cambridge, UK}
}

\author{Sugandh Pargal}
\affiliation{%
 \institution{Indian Institute of Technology Kharagpur, India}
 }
\email{sugandhpargal@gmail.com}

\author{Debasree Das}
\affiliation{%
 \institution{Indian Institute of Technology Kharagpur, India}
 }
\email{debasreedas1994@gmail.com }

\author{Tanusree Parbat}
\affiliation{%
 \institution{Indian Institute of Technology Kharagpur, India}
 }
\email{tanusree.parbat@gmail.com }

\author{Sai Shankar Kambalapalli}
\affiliation{%
 \institution{Indian Institute of Technology Kharagpur, India}
 }
\email{saishankar192@gmail.com}

\author{Bivas Mitra}
\affiliation{%
 \institution{Indian Institute of Technology Kharagpur, India}
 }
\email{bivas@cse.iitkgp.ac.in}

\author{Sandip Chakraborty}
\affiliation{%
 \institution{Indian Institute of Technology Kharagpur, India}
 }
\email{sandipchkraborty@gmail.com}

\renewcommand{\shortauthors}{Verma, et al.}

\begin{abstract}
Commuter comfort in cab rides affects driver rating as well as the reputation of ride-hailing firms like Uber/Lyft. Existing research has revealed that commuter comfort not only varies at a personalized level but also is perceived differently on different trips for the same commuter. Furthermore, there are several factors, including driving behavior and driving environment, affecting the perception of comfort. Automatically extracting the perceived comfort level of a commuter due to the impact of the driving behavior is crucial for a timely feedback to the drivers, which can help them to meet the commuter's satisfaction. In light of this, we surveyed around $200$ commuters who usually take such cab rides and obtained a set of features that impact comfort during cab rides. Following this, we develop a system \our which collects smartphone sensor data from a commuter, extracts the spatial time series feature from the data, and then computes the level of commuter comfort on a five-point scale with respect to the driving. \our uses a Hierarchical Temporal Memory model-based approach to observe anomalies in the feature distribution and then trains a Multi-task learning-based neural network model to obtain the comfort level of the commuter at a personalized level. The model also intelligently queries the commuter to add new data points to the available dataset and, in turn, improve itself over periodic training. Evaluation of \our on $30$ participants shows that the system could provide efficient comfort score with high accuracy when the driving impacts the perceived comfort. 
\end{abstract}




\maketitle

\section{Introduction} \label{introduction}
The growing success of ride-hailing services (Uber/Lyft) has increased most city dwellers' reliance across the globe on these firms, both for daily commuting and intercity travels. 
While these app-based cab services have emerged rapidly, a growing concern is the driver quality for such on-demand cab services~\cite{rogers2015social,liu2018digital}. The app-based cab companies typically employ an open business model, where both the drivers and the riders register themselves by authenticating and validating their details~\cite{ubermodel}. Typically, these cab companies continuously monitor the drivers' performance through the smartphone app's data and, more importantly, the feedback or the driver rating from the riders at the end of each trip. Such performance metrics are, in general, used for incentivizing the drivers and, therefore, are extremely important for the operational efficiency of the system. 

Although driving performance monitoring through the app-sensed data (primarily the GPS) and the riders' feedback at the end of the trip is crucial to monitor and maintain the service quality and resolve customer grievance, the current approaches have many limitations. First, the riders' feedback or the driver rating provided by the rider at the end of the trip gives only a consolidated view of their experience during the ride. It does not capture (a) the instantaneous behavior of the driving and its impact on the rider's comfort throughout the trip on a temporal scale, (b) the specific events during the trip, which have affected the riding experience. For example, a sudden jerk near the end of the trip may significantly affect the driver rating, although the rest of the trip was smooth. Indeed, various recent analysis of the Uber driver rating data has indicated that such a rating system is not accurate and also introduces multiple biases depending on the age, gender, demography, and different other factors associated with the rider as well as the riding environment~\cite{hanrahan2017roots,jiang2018ridesharing}. Second, the feedback or the rating is a complex consolidated parameter that combines multiple different factors; for example, the driver's micro-behavior towards the rider impacts the rating significantly~\cite{hanrahan2017roots}. Therefore, it lacks transparency, where the drivers and the cab companies remain unaware of the low-level factors that affected the rating for a particular ride. Although the riders may provide the reason for a low rating, it is in-general optional. As analyzed from Uber data, most riders either refrain from giving detailed feedback or share biased or random feedback~\cite{hanrahan2017roots}. Third, the impact of driving over the riding experience is very much personalized depending on the riders' age, gender, demography, health, mental conditions, etc.~\cite{rubira2016effects, verma2018comfride}. For example, although within the speed-limit, a fast-driving may cause discomfort to a commuter who is either old or physically weak but may make an office-goer happy. 

Therefore, understanding the impact of driving behavior on a personal-scale is essential for both the drivers and the app-cab companies. \revision{Considering a ride-hailing service like Uber, the smartphones of the drivers and the cab riders are typically connected through the ride-hailing service, like the Uber app. Incidentally, a cab rider's smartphone can capture her personal traits, which can also signify her comfort parameters~\cite{chittaranjan2011s}. In a collaborative environment, the rider's smartphone can continuously sense the driving data to derive the driving behavior and then correlate it with the commuter's comfort parameters. An application that understands commuter comfort could open doors for other applications like (a) a live feedback system for the driver, which provides commuter profile information and suggests what driving actions could make the commuter uncomfortable. The driver can tune or control their driving behavior based on the commuter's personal preference, making the riding more interactive and get a better rating~\cite{chan2019rating,raval2016standing}; (b) the app-cab companies can also match the drivers with the riders based on the driving profile of the driver and the riding preference of the rider.}

\textbf{Technology requirements and associated challenges:} Automated systems for generating ratings from behavioral observations can play an essential role in addressing such issues, as the works by \textit{Thebault-Spieker et al. }~\cite{10.1145/3134736}, and \textit{Liang et al. }~\cite{10.1145/2998181.2998217} have shown, either by utilizing surveys or simulations. However, to address the issues at a practical level and build various other value-added services based on the impact of driving behavior over a rider's riding experience, we need an end-to-end driving profiling toolbox. This mechanism should continuously assess the driving behavior's influence over the rider's comfort and provide critical feedback, recommendation, or alert to the driver and the cab companies. However, as we mentioned earlier, such a model should capture the riders' personality traits, as different factors have quite distinct impacts on other riders. However, these factors may not carry a direct signature to understand the impact of driving behavior on the commuter's perceived comfort. For example, on a bumpy road, even a good driver may not avoid the jerkiness altogether; therefore, the commuter's discomfort, in this case, is linked to the driving environment and not to the driving behavior. However, the driving behavior can be alarming if a sudden jerk is felt on a smooth road. Therefore, even a personalized learning model is not suitable to capture the commuter's comfort as it also widely varies across different driving environments.

Effectively, the need here is to have a model which can (a) not only take decisions at a personalized level but also take into account the differences in the road conditions or the driving environments, (b) understand different baseline signatures associated with factors like acceleration, jerk, congestion, etc. that are associated with the comfort level of the commuter under various driving environments, and (c) estimate the deviation of these signatures from their typical pattern (corresponding to commuter comfort) indicating possible discomfort for the commuter. We also target to make the profiling online, based on the streaming sensor data (accelerometer, GPS) captured from commuters' smartphones running the riding app. This ensures that our framework could be used for developing online services, such as alerting the driver during the trip itself if the passenger is likely to feel uncomfortable due to some driving actions.

\revision{In this paper, we first develop an application to collect various sensor data from the rider's smartphone while on a cab ride to link the driving style with her comfort perception (\S\ref{datacollection}). The application helps us to generate a rich dataset of driving and commuter comfort labels. Our primary contributions relying on the collected dataset are as follows;
\begin{itemize}
    \item Based on an online survey and a user study, we define what features affect a rider's comfort while on a ride (\S\ref{motivation}).
    \item We model the Spatio-temporal self-exciting (the value at the current time instance influences the value at the next time instance) features, viz. speed of the vehicle, jerkiness, and congestion, by analyzing their spatial time series distribution (\S\ref{featureextractor}).
    \item We develop a \textit{Hierarchical Temporal Memory} (HTM)~\cite{hawkins2007intelligence, ahmad2015properties} based approach to detect anomalies in the distribution of these Spatio-temporal features, which are analogous to the rider's discomfort (\S\ref{anomalydetection}).
    \item As HTM only detects anomaly for a single feature at a time; we develop a neural network model to map the likelihood of the discomfort of all the three features along with other static features to compute the rider's comfort level. Keeping personalization of comfort in mind, we opt for \textit{Multi-task Learning}~\cite{caruana1998multitask} (\S\ref{behavioranalyzer}). The model also has a feedback mechanism to improve itself with time (\S\ref{commuterfeedback}).
\end{itemize}
The developed model continuously predicts the rider's comfort level based on the driving behavior and her personality traits, and such information can be used for developing multiple applications in a driver-rider collaborative environment, as stated earlier. As a proof of concept implementation, we implement an automated driving rating system which provides continuous feedback to the driver over the ride-hailing app.}

We perform experiments over $30$ users to evaluate each block of the system. Finally, we develop a rating application based on the overall comfort felt during the trip, which uses \our as a framework (\S\ref{evaluation}). Following this, we provide a discussion of the limitations and possible future directions for this work(\S\ref{discussion}). Before proceeding into the system's details, we first give a brief survey of the related literature in the next section (\S\ref{relatedwork}).

\section{Related Work} \label{relatedwork}
In this section, we discuss related literature that built ways to define commuter comfort, a crucial unit of the transport system, and develop systems to compute comfort in different scenarios. 

\subsection{Commuter Comfort: How is it perceived?}
Understanding commuter comfort could be dated back to \textit{Mayr}~\cite{mayr_1959} coining the term \textit{traveling comfort} composed of riding, local and organizational comfort~\cite{oborne1978passenger}. Local comfort is the comfort felt on stations or airports and takes into consideration factors like comfortable transfers or condition of waiting rooms. Organizational comfort takes into consideration the comfort linked to organizational origin, like availability of transport or reliability of a service. Riding comfort, which is the comfort inside the vehicle, was later quantified by  \textit{Kottenhoff}~\cite{kottenhoff2016driving} based on the experience observed due to vehicle movements like accelerations, shaking, vibration, or jerks. So effectively, it could be linked to the driving style of the driver, which could include instances like uneven driving, heavy braking, sharp acceleration, jerkiness, as observed by \textit{Kottenhoff et al.}~\cite{kottenhoff2011samband}. In transport research literature, such as~\cite{florio2013network,shen2016analysis,tirachini2013crowding} and the references therein, personal interviews are used to measure comfort, which being time-consuming and labor-intensive lacks scalability.
Furthermore, there have been several works which have shown that comfort is a personalized concept. For instance, \textit{Clear et. al.}~\cite{clear2017d} report that in a building the perception of comfort might vary between the occupants. \textit{ComfRide}~\cite{verma2018comfride} shows that multiple factors could affect a commuter's perception of comfort in public buses, and every other commuter could give preference to a different set of features. This varies with age, sex, occupation, etc. Similarly, works like~\cite{ziadh, goel2016private} have shown similar results for commuters using taxis or ride-sharing options like Uber/Lyft.

\subsection{Participatory Sensing as a Cooperative Solution}
Advent of several participatory sensing works~\cite{10.1145/2675133.2675235, martelaro2017woz, cheng2017scalable, hossain2018active, mihoub2019wearables} opened grounds for approaches towards understanding commuter comfort from data obtained from multiple commuters. For instance, Cyclopath~\cite{10.1145/2145204.2145350} obtains \textit{bikeability} rating from multiple cyclists in a city to recommend the best route for a user. Similarly, PASSAGE~\cite{garvey2016passage} recommends safe path for pedestrians. SmartTransfer~\cite{du2018smarttransfer} provides a crowd-aware route recommendation system for public transit commuters. Works like CMS~\cite{li2013crowding}, RESen~\cite{song2012resen}, CommuniSense~\cite{santani2015communisense}, UrbanEye~\cite{verma:2016}, used commuter's smartphone sensor information to gain trip-related features. These works make use of multiple smartphone sensors like GPS, accelerometer, gyroscope, gravity sensor, etc. to obtain such information. Several of the new research works have tried to understand commuter comfort in public transport\revision{~\cite{eboli2016measuring,zhao2016evaluation,azzoug2017ridecomfort,verma2018comfride, dunlop2015tracking}}. \textit{RideComfort}~\cite{azzoug2017ridecomfort} utilizes smartphone sensors to obtain vibration-based ride comfort in train rides. \textit{Dunlop et al.}~\cite{dunlop2015tracking} used a smartphone-based survey to observe comfort perception of a commuter on a transit ride. Other works utilize smartphone sensors to get a perception of commuter comfort on buses~\cite{eboli2016measuring, chinanalysis}. 

\subsection{Commuter Comfort in Cabs}
Public transport has the privilege of fixed routes and scheduled times, the absence of which adds uncertainty to computing comfort in private cabs. There have been works that compute related factors like driving behavior, driver stress\revision{~\cite{munoz2017predicting, qi2016impact, li2018automatic,zhang2015real, nanni2016driving}} or relationship between the driver-commuter pair~\cite{10.1145/2441776.2441952}, which could indirectly impact the comfort of a commuter. \textit{Eren et al.}~\cite{eren2012estimating} utilize accelerometer, gyroscope, and magnetometer data obtained from a driver's smartphone to compute the driving behavior and estimate the commuting safety on the ride. \textit{Verma et al.}~\cite{vermapercom} utilize the roster information collected from multiple drivers to compute stress and relate that to predicting the driving behavior, which could cause possible accidents. However, these works utilize the data obtained from the driver and hence couldn't be personalized for the commuter. Works which directly target the comfort of a commuter also rely on data either from the car or the driver~\cite{ruzic2011improvement,park1998dynamic}. \textit{Join Driving}~\cite{zhao2013join} performs commuter comfort calculation using accelerometer data obtained from the driver's smartphone. \revision{A similar approach is followed by Machaj et al.~\cite{Machaj_2020} utilizing smartphone sensors.} \textit{Park et al.}~\cite{park1998dynamic} utilize vibrations observed from the commuter's body using sensors mounted on the seat to perceive comfort. On the other hand, \textit{Ruzic et al.}~\cite{ruzic2011improvement} utilize thermal sensors in the car to compute the comfort of the passenger. \textit{Elbanhawi et al.}~\cite{elbanhawi2015passenger} do look into personalized comfort for a passenger, but that is in the context of autonomous cars.

\begin{table}[]
\scriptsize
\caption{Comparing \our with existing works}
\label{comparerelwork}
{
\begin{tabular}{|l|l|l|p{1cm}|p{1.3cm}|p{1.6cm}|}
\hline
\textbf{Work} & \textbf{Sensing Method}           & \textbf{Transport Mode} & \textbf{Online/ Offline} & \textbf{Comfort Computation} & \textbf{Personalized for Commuter} \\ \hline
SmartTransfer~\cite{du2018smarttransfer}      & Public transit transaction records   & Public Buses         & Offline                 & Yes (crowd-aware only)                          & No                                \\ \hline
ComfRide~\cite{verma2018comfride}      & Commuter's smartphone sensors     & Public Buses            & Offline                 & Yes                          & Yes                                \\ \hline
Join Driving~\cite{zhao2013join}  & Driver's smartphone accelerometer & Cabs                    & Online                  & Yes                          & No                                 \\ \hline
Ruzic et al.~\cite{ruzic2011improvement}         & Cab mounted sensors               & Cabs                    & Online                  & Yes                          & No                                 \\ \hline
Elbanhawi et al.~\cite{elbanhawi2015passenger}      & Cab mounted sensors               & Autonomous Cars         & Online                  & Yes                          & Yes                                \\ \hline
\our       & Commuter's smartphone sensors     & Cabs                    & Online                  & Yes                          & Yes                                \\ \hline
\end{tabular}}
\end{table}
 
\subsection{Limitation of the Existing Works}
In a nutshell, although there exist several works on understanding the impact of driving behavior on commuter's comfort or the overall riding experience, they have the following limitations. (1) The majority of the works use offline information to understand the driving behavior and its impact on the commute experience. They cannot capture online and instantaneous impact of the driving behavior on the commute experience, and therefore is limited only to the applications for offline analysis. (2) The existing approaches fail to separate the impact of environmental factors from driving behavior. For example, \textit{Join Driving}~\cite{zhao2013join} looks into jerkiness by measuring the acceleration but does not consider whether the jerkiness is due to a bumpy road or due to a poor driving behavior. (3) The personalized preferences of the commuters based on age, gender, demography, occupation, etc. have not been captured in the existing models; therefore, the models are not suitable for providing fine-grained recommendation or alerts to the drivers.
\revision{As Table~\ref{comparerelwork} shows, \our addresses these limitations by utilizing the data from the commuter's smartphone to assess her comfort at a personal level and understand when a driving style is causing any discomfort.}

\section{Data Collection} \label{datacollection}
We developed an in-house data collection app in order to (a) conduct the pilot experiments and (b) to pre-train the models present in the developed \our system. 
The developed Android application is equipped to collect driving data from smartphone sensors and label the data based on the perceived comfort label in a 5-point scale.
This app records the inertial sensor data (accelerometer, gyroscope, magnetometer) and GPS information along with vehicle speed. Additionally, the app also takes comfort rating input from the commuters using a 5-point slider scale (1 being least discomfort and 5 being the most discomfort). The default value of this scale is set to 1. Whenever a commuter feels some discomfort, she could update a new value, which is set as the value of the comfort label until updated again by the commuter. Moreover, the app also probes the commuter in every $5 min$ of the last input to check if the label has changed. The commuter need not respond to this whence the previous label is used. The collected data is continuously streamed to a server to be uniquely stored for each commuter.

\begin{table}[!ht]
\caption{Data Collection Details}
\centering
\label{datatable}
\begin{tabular}{|c|c|c|c|c|c|c|c|}
\hline
\textbf{Participants} & \textbf{\begin{tabular}[c]{@{}c@{}}Age \\ Group\end{tabular}} & \textbf{\begin{tabular}[c]{@{}c@{}}Android \\ Version\end{tabular}} & \textbf{\begin{tabular}[c]{@{}c@{}}Total   \\ trips\end{tabular}} & \textbf{Cities} & \textbf{Days} & \textbf{\begin{tabular}[c]{@{}c@{}}Max-Min \\ Trip Length\end{tabular}} & \textbf{\begin{tabular}[c]{@{}c@{}}Max-Min \\ Trip Time\end{tabular}} \\ \hline
20                    & 20 - 50                                                       & 6.0 - 8.1                                                           & 100                                                               & 5               & 15            & 2 - 56 km                                                               & 5 - 120 mins                                                          \\ \hline
\end{tabular}
\end{table}

We distributed the developed application among 20 participants, who frequently take cabs for their commute, to collect data in a natural and uncontrolled environment. The participants were asked to start the application when boarding a cab and to stop the application on alighting. They were also asked to rate the driving anytime they felt discomfort and to update their rating anytime they felt a change in their perception of comfort. The participated commuters belonged to different age groups and used different models of smartphones like Lenovo K6 Power, Moto G5, Redmi 5, Redmi Note 5 Pro, with Android version ranging from Android 6.0 to 8.1. A brief summary of the data collection experiment has been provided in Table~\ref{datatable}. This data is used to carry a set of pilot experiments and to extract essential insights which helped us in developing the basic building blocks of \our. The details follow.  
\section{User Study: Identification of indicators of commuter discomfort} \label{motivation}

\begin{figure}[!ht]
  \centering
  \includegraphics[width=0.9\linewidth]{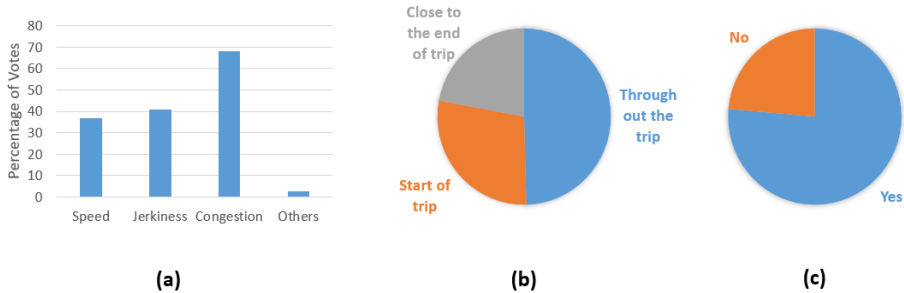}
  \caption{Online survey results. (a) Which features affect you when on a trip? (b) Which part of the trip have you usually felt discomfort? (c) Does the time of the day impact your discomfort?}
  \label{survey}
\end{figure}

First, we conduct an online survey on a set of commuters to discover the source of discomfort experienced by them in their daily commute. Next, we demonstrate the potential of those indicative features for the identification of commuter discomfort, and subsequently highlight the challenges in developing a system that can assess the comfort of the commuter leveraging on those features.

\subsection{Commuter Survey}
The objective of this survey is to identify the factors which play major role in the commuter discomfort. The survey was designed as an online Google form~\footnote{https://tinyurl.com/t93npg5} and was circulated through multiple channels like Facebook, Twitter, and WhatsApp. Additionally, it was also shared through email to different mailing groups, with which the authors were associated. The survey questionnaire is composed of multiple components. (a) First, the survey collected general information regarding the respondent, like the demography and cab usage frequency of the commuter. (b) Next it inquired the commuters about the factors, which affect their comfort when in a cab ride. Six options were provided to choose (\textit{speed, jerkiness, congestion, weather condition, driver behavior, cab condition}). These options were selected as a set of common features from existing works on riding comfort~\cite{verma2018comfride, kottenhoff2016driving, oborne1978passenger}. The commuters had the flexibility to choose multiple options. Additionally, a text box was also provided if the commuter felt any other factor should be included. (c) Furthermore, the survey queried if the discomfort she felt was usually at the beginning or end of the trip, or throughout the trip. (d) Finally, the commuter had to report if the time of the day affected the discomfort she felt on a trip. The commuter was asked to comment on the reason in a textbox for a positive response.
\subsubsection{Survey Responses}
We obtained responses from $200$ respondents who avail cab services in different cities from India, USA, Germany, Denmark, and the Netherlands. More than $70\%$ of these respondents avail cabs regularly (and $40\%$ commuters use quite frequently). The outcome of the survey has been summarized in Fig.~\ref{survey}. Majority of the respondents feel discomfort due to congestion ($67.8\%$) (Fig.~\ref{survey}(a)), followed by jerkiness ($42.5\%$) and vehicle speed ($37.7\%$). All the other factors, including user-suggested factors like \textit{cyclist/pedestrian behavior, honking by other vehicles}, collectively received $~5\%$ responses. The responses also showed that more than $50\%$ of the commuters face discomfort either at the start or the end of the trip (Fig.~\ref{survey}(b)). Furthermore, the time of the day also affects the discomfort of a majority of the commuters ($74.7\%$) (Fig.~\ref{survey}(c)). This discomfort induced from the trip time has been attributed by their illustrative responses, such as \textit{"poor driving at late night is more dangerous and hence uncomfortable than in the day"} or \textit{"it's possible to miss potholes or bumps at night which causes more discomfort"}.

\subsubsection{Lessons learnt}
Our survey study reveals that (i) \textit{speed of the vehicle}, (ii) \textit{jerkiness}, and (iii) \textit{road congestion} are the key indicators for assessing the commuter discomfort. Additionally, the segment of a trip, which causes discomfort, can be characterized as (iv) \textit{ time spent on the trip (travel time)} and (v) \textit{distance covered on the trip (distance travelled)}. Moreover, (vi) \textit{time of the day}, when the commuter is taking the trip would also be an important feature to predict commuter discomfort.

\begin{figure}[!ht]
  \centering
  \includegraphics[width=1\linewidth]{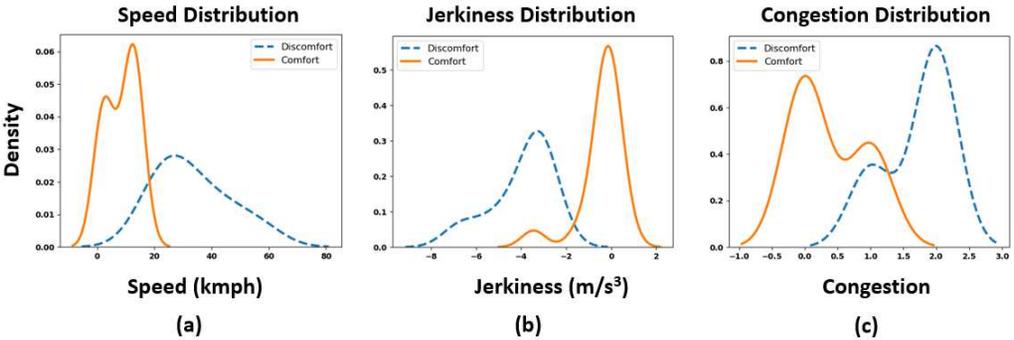}
  \caption{Kernel Density Estimate plot (with Gaussian kernel) of all the three features for samples of comfort and discomfort instances on a single trip. (a) Speed (b)Jerkiness (c) Congestion}
  \label{featuredistrib}
\end{figure}

\subsection{Opportunities and Challenges}
We conduct a pilot study to show the potential of the aforesaid indicators in order to discriminate comfortable vs uncomfortable ride of a commuter. We collect the recorded data obtained from the pilot data collection experiment on 20 participants (see Sec~\ref{datacollection}) and extract the key features (i) \textit{speed of the vehicle}, (ii) \textit{jerkiness} from acceleration data~\cite{nygaard1999method}, and (iii) \textit{road congestion} from the inertial sensors (accelerometer, gyroscope, magnetometer) following the standard techniques in~\cite{sigverma:2016}.
We consider the comfort labels 1-3 as \emph{comfortable} and 4-5 as \emph{uncomfortable} for this experiment. In Fig.~\ref{featuredistrib}, we plot the kernel density estimate, with a Gaussian kernel, of these three features for instances when the commuter is in a comfortable and uncomfortable state on the same trip. It is interesting to observe that the distribution varies considerably for discomfort state as compared to comfort for all the three features. This points to the fact that just by observing any kind of variation in the distribution of the features (\textit{speed of the vehicle}, (\textit{jerkiness}, and \textit{road congestion}), one can automatically perceive once the commuter starts feeling uncomfortable in a ride. 

\revision{The above observation provides us with two possible approaches. First, to develop a model solely dependent on Machine Learning, which would generate a trained model learned from a large dataset of commuter data. The second approach could be to estimate the features of the commuter's comfort perception, any variation (or anomaly) observed over this estimate can be perceived as commuter discomfort. However, any learning based approach would require a large volume of data and the learning would be historical. On the contrary, the anomaly detection approach, as we show later, has two advantages. First, it could work with sparse data resolving the need of a large dataset. Second, it performs learning after a trip starts with a small duration of bootstrapping at the beginning of the trip which provides an option for online learning.}

However, to leverage on the aforementioned opportunity, we need to address the following challenges.
(a) Notably, unlike, travel time, distance, and time of the day, which can be directly calculated at any time instance, the other three features (\textit{speed of the vehicle}, \textit{jerkiness}, and \textit{congestion}) vary spatially as well as temporally. Hence, it is non-trivial to estimate these features directly at any time instance; rather their values can be estimated from the modeled distribution of the features. The \textbf{first challenge} roots out from this need to develop suitable spatio-temporal baseline models, which can represent those features at the comfort state of the commuter in a ride.
(b) Subsequently, any deviation (or anomaly) from the modeled baseline distribution (termed as \emph{comfort distribution}) of features can be identified as discomfort. Hence, in every new trip, such anomaly likelihood needs to be learned for each feature, starting at the beginning of the trip. Moreover, since the learning would start at the beginning of the trip, the data available would be quite sparse. Hence the \textbf{second challenge} would be to develop a model for detecting anomaly from the \emph{comfort distribution}, that can learn well on sparse data too.
(c) The \textbf{third challenge} arises from the understanding that each commuter is different, and her personality traits should be addressed while designing the models.
(d) The performance of the pre-trained model starts deteriorating once (i) the commuter's personal preferences change over time, (ii) a new commuter launches the system. In both the cases, the pre-trained model fails to capture the comfort distribution and anomaly likelihood. Hence, the \textbf{fourth challenge} is to update and adapt the system with suitable model retraining mechanism.


\begin{figure}[!ht]
  \centering
  \includegraphics[width=1\linewidth]{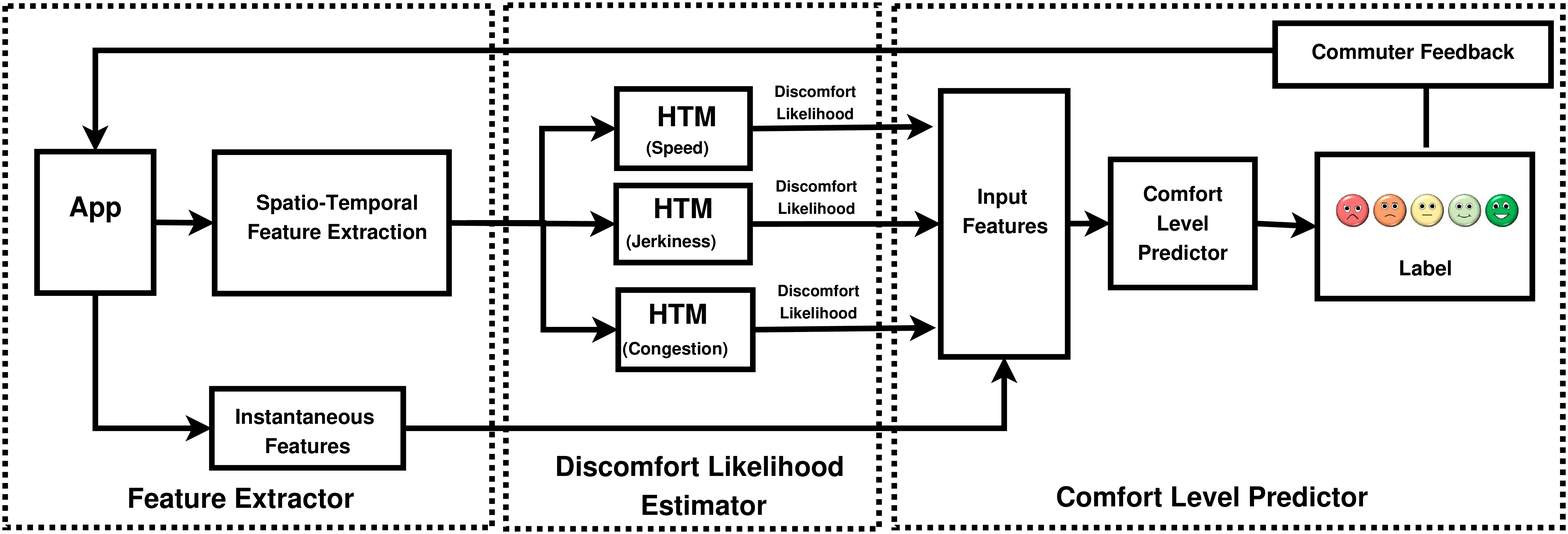}
  \caption{Block diagram of the developed system}
  \label{system}
\end{figure}


Keeping the above challenges in mind, we develop \our, which is composed of three broad modules, as shown in Fig. \ref{system}; (a) \textit{Feature Extractor} - which takes care of sensing data and extracting required features, (b) \textit{Discomfort Likelihood Estimator} - which estimates the likelihood that the driving could cause discomfort, and (c) \textit{Comfort Level Predictor} - which based on the discomfort likelihood predicts the comfort level of the commuter. \our runs a smartphone app which captures the data and finally displays the feedback, and the overall processing of the system (primarily the above three modules) run on a server. The smartphone app periodically sends the collected data to the server and fetches the feedback to display it over the app.

In the following sections we describe each module in detail.

\section{Feature Extraction} \label{featureextractor}
Taking clues from the user study (Section~\ref{motivation}), in this section we introduce the features, which carry the signature of commuter comfort. We rely on the sensor data streams collected from the smartphones to extract those features.



As smartphone sensor data are usually noisy, we perform pre-processing using standard techniques for axis reorientation and data smoothing~\cite{verma:2016}. Following this, we concentrate on the extraction of features, which can be broadly categorized into two classes (a) \textit{instantaneous} and (b) \textit{spatio-temporal features}.

\subsection{Instantaneous Features}
These features could be calculated directly from the sensor data at any time instance. From the commuter survey, we identify three instantaneous features, namely (a) travel time ($T_t$), (b) distance traveled ($d_t$), and (c) time of the day ($Z$), which may directly impact the discomfort of a commuter. $Z$ is divided into 4 time zones (6 AM -10 AM(0), 10 AM- 4 PM(1), 4 PM - 10 PM(2), 10 PM - 6 AM(3)) in this paper, however it is configurable and would change based on the city characteristics.

\subsection{Spatio-Temporal Features}
Unlike instantaneous features, these features vary both spatially as well as temporally, hence are difficult to compute at any time instance. For instance, determining the exact \emph{speed} of a vehicle at any point is difficult, as it depends on both time and the spatial characteristics of the road the vehicle is driving on. In our survey, we identify three spatio-temporal features, (a) speed ($v$), (b) jerkiness ($j$), and (c) congestion ($c$). The exact values of these features depend on the actual time \& location of the vehicle as well as the behavior of neighboring vehicles at the time of computation. Instantaneous values of these spatio-temporal features do not directly indicate the commuter discomfort. Rather, at any point of journey, we may estimate the instantaneous value of the spatio-temporal feature at that time \& location and then compute the discomfort likelihood, based on the deviation of the feature values from their baseline (comfort) distribution as perceived by the commuters in the previous trips (details in Sec 7).

The instantaneous \emph{speed ($v$)} can be obtained from the GPS sensor. The instantaneous value of \emph{jerk ($j$)} is computed as $da(t)/dt$ within a sampling window of $5s$~\cite{nygaard1999method}, where $a(t)$ is the acceleration along y-axis at time $t$. On the other hand, the instantaneous value of \emph{congestion (c)} could be calculated by observing the stop-move-stop-move pattern of acceleration along y-axis~\cite{verma2017smart}. Let the time period for the stop-move pattern be $t_{sm}$, then we have medium congestion (1) when $ 1min \leq t_{sm} < 5min$ and high congestion (2) when $ t_{sm} \geq 5min$. Otherwise, the congestion value is set to zero.

\revision{It is interesting to note that although the features like congestion and time of the day cannot be explicitly controlled by the cab driver, nevertheless, this is important to observe how the cab driver deals with such scenarios; this discriminates between efficient driving with the poor driving and impacts the (relative) comfort of the commuter. By modeling the comfort distribution for congestion (as spatio-temporal feature), we speculate the driving behavior, which provides (relative) comfort to the commuter in congestion.}

\section{Discomfort signature from the spatio-temporal features}\label{anomalydetection}
Notably, detecting commuter comfort from spatio-temporal features is not trivial; the instantaneous values of the spatio-temporal features would not directly provide a measure of comfort. For example, on a bumpy road, the jerkiness is likely to be higher -- even an expert driver cannot avoid that completely. However, in this case, although the commuter may feel discomfort, it is not due to the driving behavior, rather due to the driving environment. Even a personalized model does not help, as the trip environment, like road condition, congestion, etc. may vary for each trip, which may affect the commuter comfort.

In order to address this issue, we develop a model which could identify the commuter's discomfort at a personalized level on any trip. This model has two important steps. First, we model the baseline distribution of the spatio-temporal features perceived at the comfort state of the commuter; we call these distributions as \textit{comfort distribution}. Importantly, the \textit{comfort distribution} would exhibit different behavior on different trips. For instance, the distribution of \emph{jerkiness} on a bumpy road would be different compared to that on a smooth highway. In the second step, we aim to estimate the spatio-temporal features from the extracted sensor data at any point of time of the trip, observe its deviation from the baseline \textit{comfort distribution} and compute the likelihood of commuter discomfort. Hence, we train the \textit{comfort distribution} to a learning model which can compute the deviation of the estimated distribution of the spatio-temporal features from the comfort distribution for the commuter on that trip.  We designate this deviation as the \textit{discomfort likelihood}. The detail follows.


\subsection{Step 1: Modeling \emph{Comfort Distribution} of Spatio-Temporal Features}
We now focus on modeling the distribution of speed, jerkiness, and congestion at the comfort state of the commuter, which are represented as spatial time series. We start with the \textbf{speed} ($v$) which at any time instance could take any random value in a metric space; however, it is always dependent on the time instance and occurs over the period $[0, T]$, where $T$ is the total trip time of the commuter. Moreover, past speed history impacts the current speed of the vehicle. Consequently, we model $v$ as a self-exciting temporal point process~\cite{reinhart2018review}. Hawkes proposed the concept of a self-exciting temporal point process~\cite{hawkes1971spectra} based on the notion of causality, i.e., if an event occurs, another event becomes more likely to occur locally in time. If $\mathcal{H}_t$ is the history of all speed events in a trip, for which the commuter felt comfortable up to time $t$, then the \textit{conditional intensity}~\cite{rasmussen2011temporal}, which characterizes the speed process is represented as;
\begin{equation}\label{setpp}
\lambda_v(t|\mathcal{H}_t) = \mu(t) + \int_{0}^{t} g(t - u) \texttt{d}v(u)
\end{equation}
Here, $\mu$ is \emph{background rate} of the speed events describing how the likelihood of the speed values evolves in time. $g(t)$ is called \textit{triggering kernel}, which regulates the influence of recent history vs. older history on the current value of speed~\cite{hawkes1971spectra}.

Next we turn toward the other two features \textbf{jerkiness} ($j$) and \textbf{congestion} ($c$). Unlike speed, both of these features are affected by the time and spatial information. For instance, congestion observed by a vehicle at some location is obviously regulated by the current time $t$. Nevertheless, the congestion felt by that vehicle at time $t$ also gets affected by the action of nearby vehicles present in that location, attributing the role of spatial factor. Similarly, the jerkiness of the vehicle is impacted by the spatial characteristics of the road.
We suitably extend the temporal point process of Eq.~\ref{setpp} to model the jerkiness and congestion as a self-exciting spatio-temporal point process~\cite{reinhart2018review}.
The conditional intensity function which characterizes a spatio-temporal self-exciting process for feature ($f$), where $f$ could be $j$ or $c$ at times $t \epsilon (0,T]$, and at locations $r \epsilon X \subseteq \mathbb{R}^d$ can be expressed as;
\begin{equation}\label{sestpp1}
\lambda_f(r,t|\mathcal{H}_t) = \mu(r,t) + \int_{0}^{r}\int_{0}^{t} g(r - u, t - w) \texttt{d}f(u) \texttt{d}f(w)
\end{equation}


\subsection{Step 2: Discomfort Estimation from the Spatio-Temporal Features}

In this section, we compute \textit{discomfort likelihood} as the deviation of the \emph{observed distribution} of the spatio-temporal features during a trip, with respect to the modeled \emph{comfort distribution}. During a trip, we estimate the observed spatio-temporal features, \textit{speed $v$}, \textit{jerkiness $j$}, and \textit{congestion $c$} from the recorded smartphone sensor stream following Sec 6.2. We develop a HTM based model which we first train on the \textit{comfort distribution}. Hence, this HTM model allows us to predict the spatio-temporal feature, pretending the comfortable state of the commuter. On the run time, during a trip, the model computes the deviation of this predicted and observed features as \textit{discomfort likelihood}, indicating the commuter discomfort. Fig 4 summarises the procedure.

\subsubsection{Predicting spatio-temporal features from \emph{comfort distribution}}
We develop a Hierarchical Temporal Memory (HTM)~\cite{hawkins2007intelligence, ahmad2015properties} model (see Fig 4) to predict the spatio-temporal features $\pi(x_t)$ at time $t$, pretending the comfort state of the commuter. First we encode the instantaneous value of a spatio-temporal feature as input $x_t$ semantically as a sparse array called the \textit{Sparse Distributed Representation (SDR)} through a \textit{spatial pooler} to get $a(x_t)$. Then using the \emph{comfort distribution} for each of the three spatio-temporal features, obtained from Eq.~\ref{setpp} and ~\ref{sestpp1}, we train the HTM model in temporal pooler such that the predicted $\pi(x_t)$ is equal to $a(x_t)$. In this way, the HTM model gets trained to predict any spatio-temporal feature (from the \emph{comfort distribution}) at a given time $t$.

\begin{figure}[!ht]
  \centering
  \includegraphics[width=0.7\linewidth]{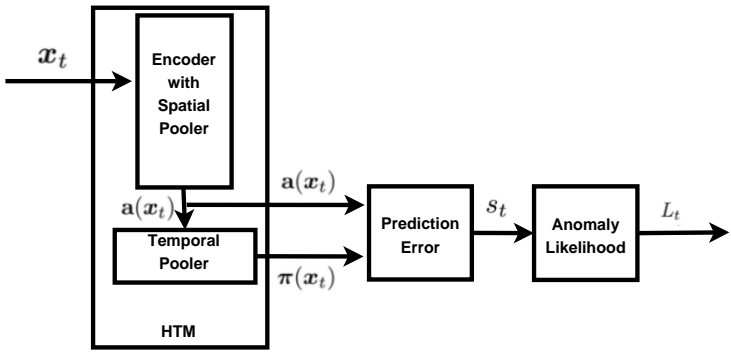}
  \caption{HTM Architecture and Anomaly Detection}
  \label{htm}
\end{figure}

\subsubsection{Estimating anomaly: deviation of observed and predicted features:} 
Hierarchical Temporal Memory (HTM) model is equipped to detect the anomaly observed in the \emph{sparse data} obtained from the commuter's smartphone. At run time during a trip, the instantaneous value $x_t$ of the observed spatio-temporal feature is fed as input to the trained HTM model, which is then again represented as $a{x_t}$ using the encoder. The temporal pooler, on the other hand, predicts the expected value $\pi(x_t)$ at the comfortable state of the commuter. 
Given the observed representation $a(x_t)$ and the predicted representation $\pi(x_t)$ of the current feature input $x_t$, the prediction error is computed
that will be 0 for accurate prediction and 1 for completely orthogonal prediction.

\subsubsection{Discomfort Likelihood Calculation:} \label{al1}
Notably, prediction error only shows the instantaneous predictability of the system. For instance, a sudden brake may or may not lead to poor driving. Thus, a threshold on the prediction error would not be a proper measure of commuter discomfort. Rather, HTM model relies on the distribution of errors as a discomfort metric. It stores a window of last $W$ prediction errors as \textit{raw anomaly scores} and models the distribution as a rolling-normal distribution. 
Given the sample mean $\rho_t$ and variance $\sigma_t$ in $W$, HTM then calculates a recent short-term average of the raw anomaly scores, and computes the \emph{discomfort likelihood} based on the Gaussian tail probability (Q-function)~\cite{karagiannidis2007improved}.
\begin{equation}
\mathcal{L}_t = 1 - \mathcal{Q} \left( \frac{\widehat{\rho_t} - \rho_t}{\sigma_t} \right)
\end{equation}
where $\widehat{\rho_t}$ is the sample mean for the short-term moving average window $W'$, where $W' << W$.

We calculate this likelihood score for all the three spatio-temporal features, (i) speed of the vehicle ($v$), (ii) jerkiness observed ($j$), and (iii) congestion on the road, from the recorded smartphone sensor data during the trip of a commuter.

\section{Development of \our} \label{behavioranalyzer}
In this section, we develop \our which infers the comfort level of a commuter, based on the driving quality during a trip. The core of \our is a Multi-task Learning (MTL)  model, which leverages on the \textit{discomfort likelihood} of spatio-temporal features along with the instantaneous features, to indicate the commuter comfort on a 5-point scale. \our captures the personality traits of the individual commuter as well as adapts $\&$ retrains itself for a newly joined commuter or if the existing commuter changes her preferences. \revision{It is interesting to note that, in principle, the proposed MTL model for commuter comfort detection may indeed work on the raw data. However, the MTL model requires a massive volume of raw data to automatically learn the complex Spatio-temporal features, as stated before, and it should take a long time for loss convergence.  As we wish to develop a personalized MTL model equipped to predict the rider's comfort in real-time on a trip, the availability of a sufficient volume of data on a trip is a significant challenge.  This challenge gets manifold if we allow MTL to automatically learn those complex features from raw data; fast convergence of loss is another issue. Hence, in our data constraint environment, we handcraft those complex features (as discomfort likelihood of spatial features) and feed them to the MTL model, which allows us to `quickly' train the model with `reasonably sparse volume' of data. }


\subsection{MTL driven comfort detection}
We develop the model to identify the commuter's perception of driving using the Multi-task Learning technique~\cite{caruana1998multitask}. The perception of each commuter is taken as a separate task, thus taking into consideration the personality trait of the commuter regarding the driver's driving style. Additionally, the model also ensures robust learning by sharing the data across multiple tasks to learn features of one commuter (one task) from related commuters. The model provides an \emph{indicator vector} of dimension 5 (for 5-point comfort scale), designating a probability for each comfort level (ranging from completely comfortable (1) to highly uncomfortable (5)). The comfort level with the highest probability gets inferred as the perceived comfort of the commuter.

\begin{figure}[!ht]
  \centering
  \includegraphics[width=0.6\linewidth]{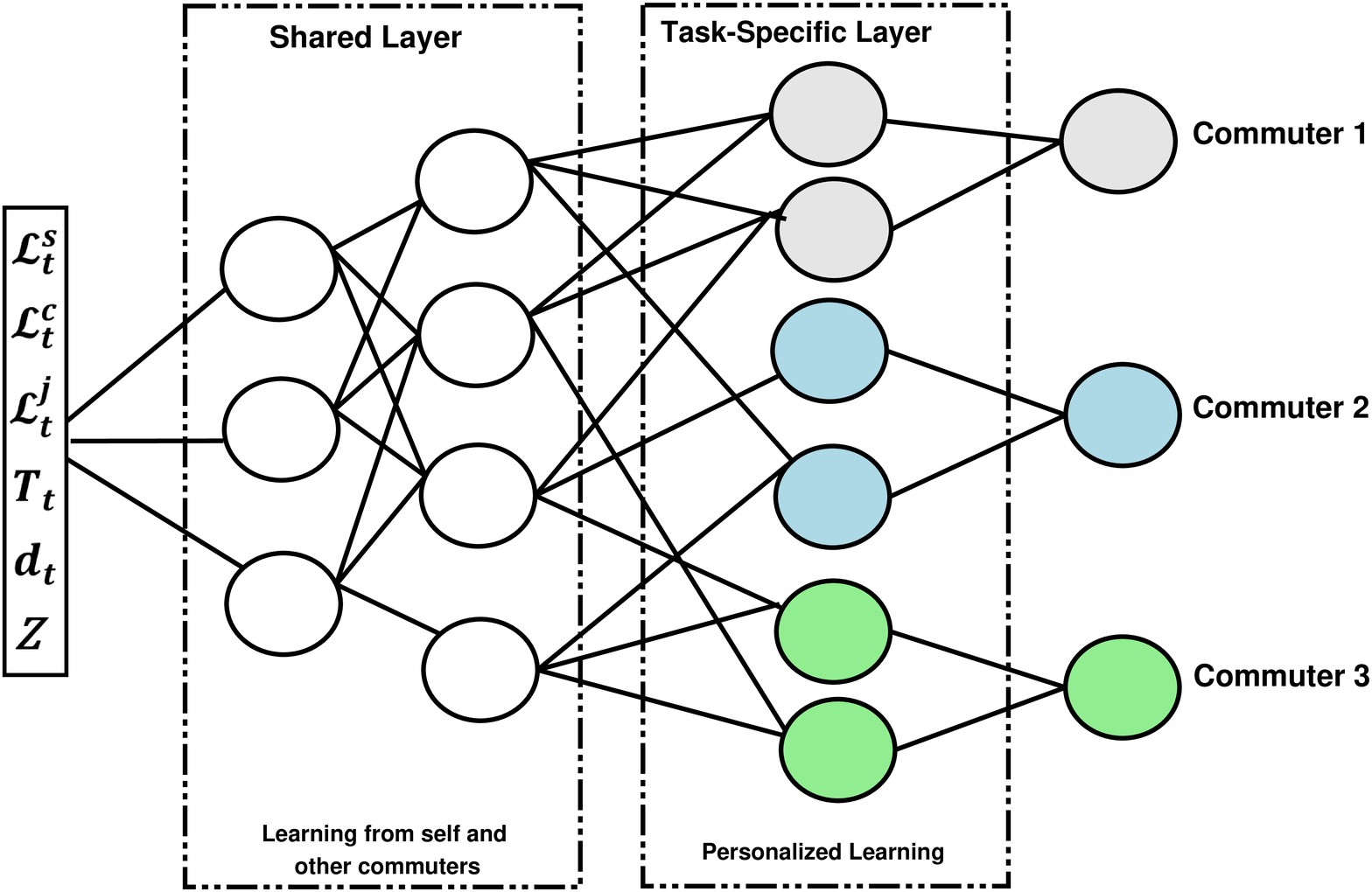}
  \caption{Architecture of an MTL-NN}
  \label{mtl}
\end{figure}

Effectively, as shown in Fig.~\ref{mtl}, the \textit{Multi-task learning Neural Network} (MTL-NN) model learns the features at two levels, the \textit{shared} and the \textit{task-specific} levels. The input containing the feature vector, obtained from the spatio-temporal and instantaneous features, which it obtained from the previous layer, is fed into the model. The next layer is the \textit{shared layer}, which contains a set of hidden nodes; the parameters of these nodes are shared across other nodes of this layer for all the tasks. This shared layer enables inductive transfer which improves learning for one task (say, the impact of congestion on one commuter) by using the information contained in the training signals of other related tasks (say, the impact of congestion on other commuters who are similar to the commuter in some way). This improves the overall model performance since some features may be easy to learn for \textit{Commuter 1} while being difficult to learn for \textit{Commuter 2}. This might occur since \textit{Commuter 2} interacts with those features in a more complex way than \textit{Commuter 1}. The shared layer allows the model to eavesdrop from \textit{Commuter 1}, and learn the features for \textit{Commuter 2}.

Simultaneously, MTL-NN allows few hidden nodes to become specialized for capturing the comfort perception of just one commuter (i.e., specialized in one task); this personalized computation, capturing the characteristics of the specific commuter, is carried out in the final \textit{task-specific layer}. In this layer, computation of one specific commuter can ignore the hidden nodes connected to other commuters, by keeping the weights connected to the small, as they do not appear useful. In this layer, the learning mechanism maps the generalized information learned at the \textit{shared layers} to a final prediction personalized by the characteristics of the specific commuter (task).

\subsection{Model adaptation and retraining} \label{commuterfeedback}
We initially train the \our based on the pilot data collected in Section 3. However, \our is equipped to adapt itself, once the performance of the model drops significantly. Precisely, as the confidence of comfort prediction deteriorates, the model occasionally probes the commuter for ground truth comfort levels (without resulting in survey fatigue). The drop in the prediction confidence is determined from the probabilities in the indicator vector; comparable probability values across diverse comfort levels (say, level $2$ and $5$) in the vector indicate the compromise in the prediction quality\footnote{In our implementation, we set a difference of $0.1$ between the highest and next highest probability in the indicator vector as the threshold for probing.}. Subsequently, the commuter responses are uploaded on the server to retrain the MTL-NN model with newly collected labeled data. This facilitates \our to enrich the dataset with more data points, both from existing and newly joined commuters and, in turn, improves the model by training on a higher volume of data.

\section{Evaluation} \label{evaluation}
We followed a client-server model for implementing \our. The server takes care of the major computation tasks like feature extraction, discomfort likelihood computation, and comfort level calculation while the client handles the data collection, shows the computed driving feedback details to the commuters, and logs the commuter responses about their feedback on the driving. The \textit{Discomfort Likelihood Computation Model} and the \textit{Comfort Level Predictor} are both written in Python over a Debian 9.3 server, with an Intel(R) Xeon(R) E5-2620 v3 @ 2.40GHz CPU, 32GB memory and 16GB GPU. The client is built for Android and was published on the Google Play Store. We performed a measurement study of the app using the \textit{Android Profiler Toolkit} on three devices, \revision{Lenovo K6 Power (Android v6.0.1, API23, sampling rate of 3$\mu$s), Moto G5 (Android v7, API25, sampling rate of 3$\mu$s), and Samsung J8 (Android v8, API26, sampling rate of 3$\mu$s)}. The measurement study shows that the application utilizes $\approx20\% (\pm3\%)$ CPU resources and $\approx95MB (\pm5MB)$ memory on an average in an hour. Battery consumption was $5\%$ on an average over an hour of the total battery consumption, and the energy consumption was \textit{light} overall. 
%

\begin{figure}[!ht]
  \centering
  \includegraphics[width=0.6\linewidth]{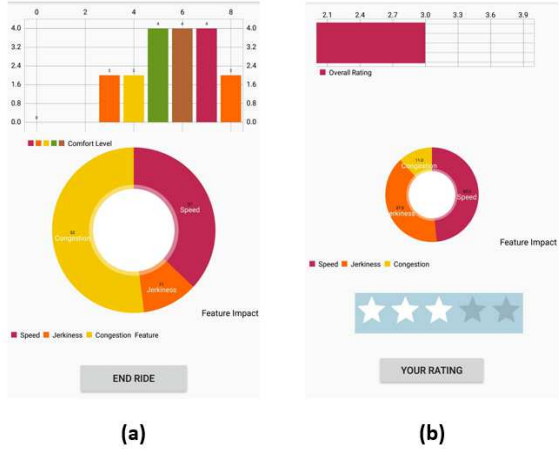}
  \caption{App UI (a) App interface showing comfort levels at intervals and impact of speed, congestion and jerkiness over the comfort. (b) The overall rating unit of the app. (1 - Most Comfortable, 5 - Most Uncomfortable)}
  \label{app}
\end{figure}

Fig~\ref{app}(a) shows the UI of the app, indicating the projected comfort level of the commuter and the percentage impact of speed, congestion, and jerkiness over the comfort level. The app is available on Google Play Store and had $30$ users at the time of reporting\footnote{https://play.google.com/store/apps/details?id=com.rohit.ridecomfort} with $5$ of these also having taken part in the data collection experiment for pilot study and model training (as mentioned in Section~\ref{datacollection}). The data from the remaining $25$ users have not been used for model training and has been used entirely to test the performance of \our. The users were advised to install the application in their smartphone and \revision{start the application every time they took a cab ride. Once started the app could be sent to the background.} The commuters were also requested to provide proper feedback whenever the system queried them. The smartphone sampled data in a window of $5 seconds$, a threshold as per literature for jerkiness calculation~\cite{nygaard1999method}, and thus used for all other features. If not connected to the Internet, the information was stored in a temporary file and uploaded to the server when the smartphone got reconnected. For the experiment phase, the users were also asked to run the data collection application to obtain the ground truth labels. \revision{The dataset details are given in Table~\ref{dstats}.}


\begin{table}[!ht]
\hfill
\centering
\caption{\revision{Dataset Statistics}}
\label{dstats}
{
\begin{tabular}{|l|p{1cm}|p{1cm}|p{1cm}|p{1cm}|p{1cm}|l|}
\hline
\multirow{2}{*}{\textbf{Count}} & \multicolumn{5}{c|}{\textbf{Percentage of Count for each rating}} & \multirow{2}{*}{\textbf{Mean}} \\ \cline{2-6}
                                & \textbf{1}  & \textbf{2}  & \textbf{3}  & \textbf{4} & \textbf{5} &                                \\ \hline
7431376                         & 2           & 32          & 33          & 32         & 1          & 3                              \\ \hline
\end{tabular}}
\end{table}

In this section, we first provide the evaluation of the complete system compared to other existing systems. Following this, we look into the performance of the \textit{Discomfort Likelihood Estimator} sub-module, followed by the \textit{Comfort Level Predictor}. Finally, we provide a use case of the complete system by showing its usage in a driver-rating application.

\subsection{Competing Systems}
We compare \our with two competing systems which also provide commuter comfort level using smartphone data. As there are not many such systems available, we compare with one system which is developed for buses~\cite{chinanalysis} (could be easily extended to cabs) and the other for cars~\cite{zhao2013join}. Both of these models could be used for online comfort level computation. Additionally, we also develop another model which is similar to \our but gets trained over each commuter in isolation.

\subsubsection{\textit{Chin et al.}~\cite{chinanalysis}} 
This work provides a method which utilizes statistical analysis using classification and regression tree method to compute commuter comfort. They utilize kinematic data collected from commuter smartphone and label the comfort into three levels (\textit{No discomfort (1)}, \textit{Noticeable discomfort (2)} and \textit{Annoying discomfort (3)}). We implemented the model on the available data and generated a pruned tree of size $26$. The terminal nodes were then labeled for the three comfort labels. As \our is on a 5-point scale, we use the standard Likert Scale relabeling strategy~\cite{relabel}; using integral labels, we obtain the mapping as, $(1,2) \rightarrow 1; 3 \rightarrow 2; (4,5) \rightarrow 3$.

\subsubsection{Join Driving~\cite{zhao2013join}}
\textit{Join Driving} gets the commuter comfort from acceleration data only on a 6-point scale utilizing the  International Standard 2631-1-1997~\cite{iso1997mechanical}. They compute the vibration felt by the commuter using the total value of weighted root mean squared acceleration, combining the vibration along all axes as $a_v = \sqrt{(1.4a_x)^2 + (1.4a_y)^2 + (a_z)^2}$. Again using the relabeling strategy, we obtain the labels on a 3-point scale as, $(1,2) \rightarrow 1; (3,4) \rightarrow 2; (5,6) \rightarrow 3$.

\subsubsection{Single Task Learning}
In the Single Task Learning (STL) approach, we use the same architecture of \our, while replacing the MTL-NN with an STL-NN. Thus, the model has to learn over each commuter in isolation whenever a new commuter installs the application. Here also, we perform the mapping to a 3-point scale as in~\cite{chinanalysis}.

\begin{table}[!ht]
\centering
\caption{AUC values for the Competing Systems}
\begin{tabular}{l|l|l|l|l|}
\cline{2-5}
                                  & \multicolumn{1}{c|}{\textbf{1}} & \multicolumn{1}{c|}{\textbf{2}} & \multicolumn{1}{c|}{\textbf{3}} & \multicolumn{1}{c|}{\textbf{Avg}} \\ \hline
\multicolumn{1}{|l|}{\textbf{\our}}  & 0.87                            & 0.86                            & 0.889                           & 0.873                             \\ \hline
\multicolumn{1}{|l|}{\textbf{STL}} & 0.73                            & 0.753                           & 0.726                           & 0.712                             \\ \hline
\multicolumn{1}{|l|}{\textbf{~\cite{chinanalysis}}} & 0.71                            & 0.721                           & 0.68                            & 0.704                             \\ \hline
\multicolumn{1}{|l|}{\textbf{~\cite{zhao2013join}}} & 0.53                            & 0.658                           & 0.547                           & 0.578                             \\ \hline
\end{tabular}
\label{competecomfort}
\end{table}

We ran the three models along with \our over multiple trips. In each of these trips, the comfort level provided by each of these models were stored simultaneously. However, as~\cite{chinanalysis} had only three levels, we map the comfort levels on a 3-point scale. Table~\ref{competecomfort} shows the AUC values for the individual comfort levels and also the average AUC. We take into account factors like congestion or time of the day in addition to the kinematic data to compute comfort, which helps in improving the result when compared to~\cite{chinanalysis} and~\cite{zhao2013join}. The shared learning and personalization aspect of the model helps \our to get an edge over the STL model, as STL based models couldn't capture the personality traits as MTL could~\cite{caruana1998multitask}. This personalization aspect also is a shortcoming with the other two approaches.

\subsection{Performance Evaluation: Discomfort Likelihood Estimator}
\our doesn't exactly give an explicit result showing if a distribution is anomalous. Instead, it uses the discomfort likelihood ($\mathcal{L}_t$) to analyze the driving behavior. In order to evaluate the HTM module, we threshold this $\mathcal{L}_t$ over a configurable parameter $\epsilon$. We say an anomaly is detected if $\mathcal{L}_t \geq (1 - \epsilon)$~\cite{ahmad2017unsupervised}. Usually, the standard value of $\epsilon$ is used as $10^{-5}$, which we used for our model also. We set $W$ and $W'$ (described in Section~\ref{al1}) as $4000$ and $10$, respectively, which we obtained empirically as shown in Table~\ref{ww}. We compare the HTM model with two competing models -- (a) Multinomial Relative Entropy~\cite{wang2011statistical} and (b) EXPected Similarity Estimation (EXPoSE)~\cite{schneider2016expected}. These models which are state-of-the-art anomaly detection models were selected keeping in mind that the algorithms should; (a) make online predictions, (b) learn continuously and in an unsupervised fashion, (c) adapt to dynamic environment changes and (d) should make anomaly detection as early as possible. Both these algorithms have open-source implementation~\footnote{\url{https://github.com/numenta/NAB/tree/master/nab/detectors} (Access: \today)}. We performed parameter tuning empirically and set the thresholds at our end, as mentioned below. These were kept fixed across all streams of data.

\subsubsection{Multinomial Relative Entropy (RE)~\cite{wang2011statistical} }
This algorithm compares the observed data against multiple null hypotheses while representing frequencies of quantized data over certain window sizes. 
In the implementation, we tuned the window size and the bin count, which were set as $55$ and $10$, respectively. The chi threshold, which is used to determine if a hypothesis has occurred frequently, was set as $1$.

\subsubsection{EXPoSE~\cite{schneider2016expected}}
The EXPected Similarity Estimation (EXPoSE) approach is based on the likelihood of the current data-point being normal based on the inner product of its feature map with kernel Hilbert space embedding of the older data points with no assumption of the underlying data distribution. 
We have used the decay variant of EXPoSE, which provides better results compared to windowing~\cite{schneider2016expected}. Here we tuned the decay factor to be set as $0.01$.

\begin{figure}[!ht]
\begin{minipage}{0.48\linewidth}
  \centering
  \includegraphics[width=\linewidth]{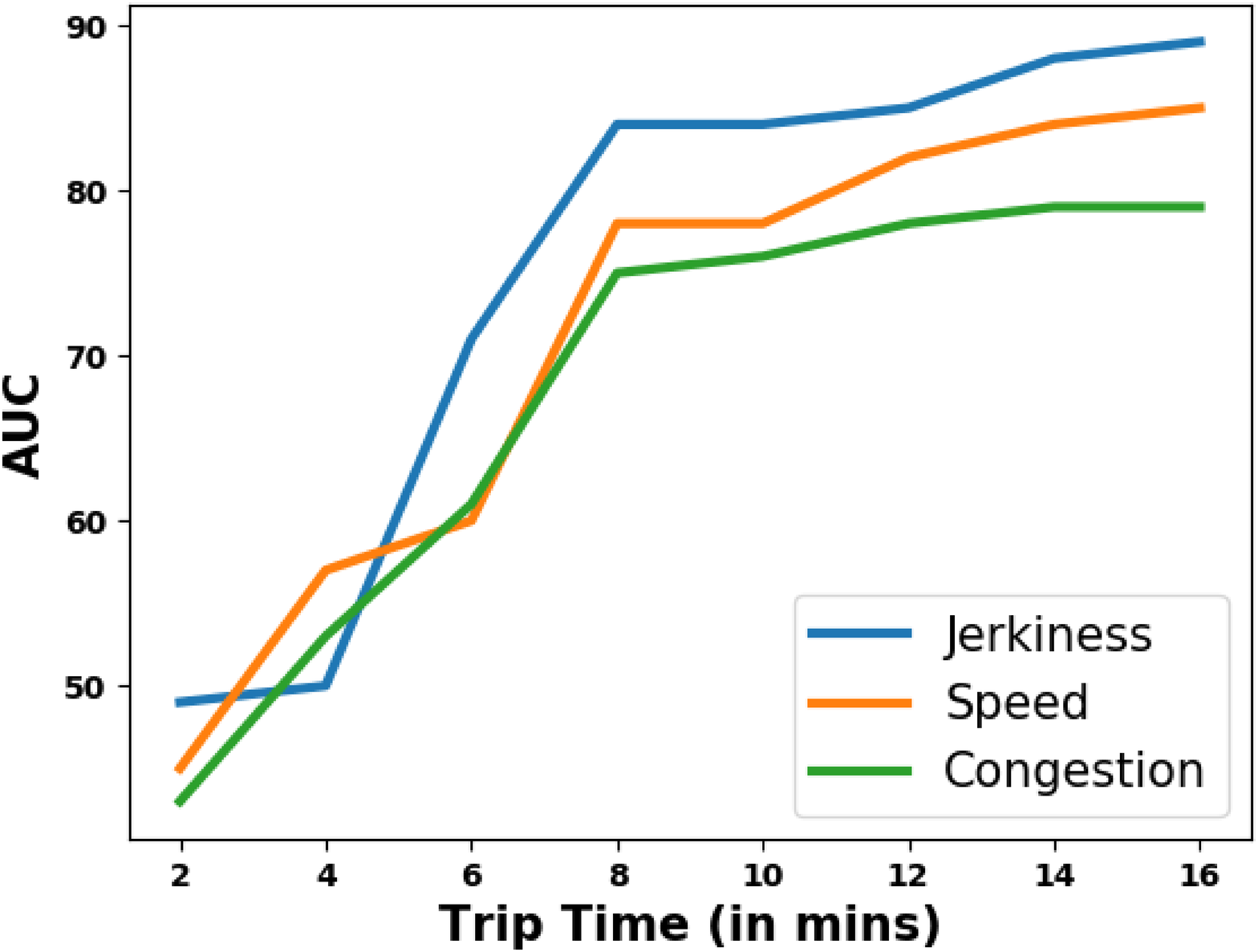}
  \caption{Change in AUC with percentage of data used for training for all the three features.}
  \label{htmmin}
\end{minipage}
\hfill 
\begin{minipage}{0.48\linewidth}
  \centering
  \includegraphics[width=\linewidth]{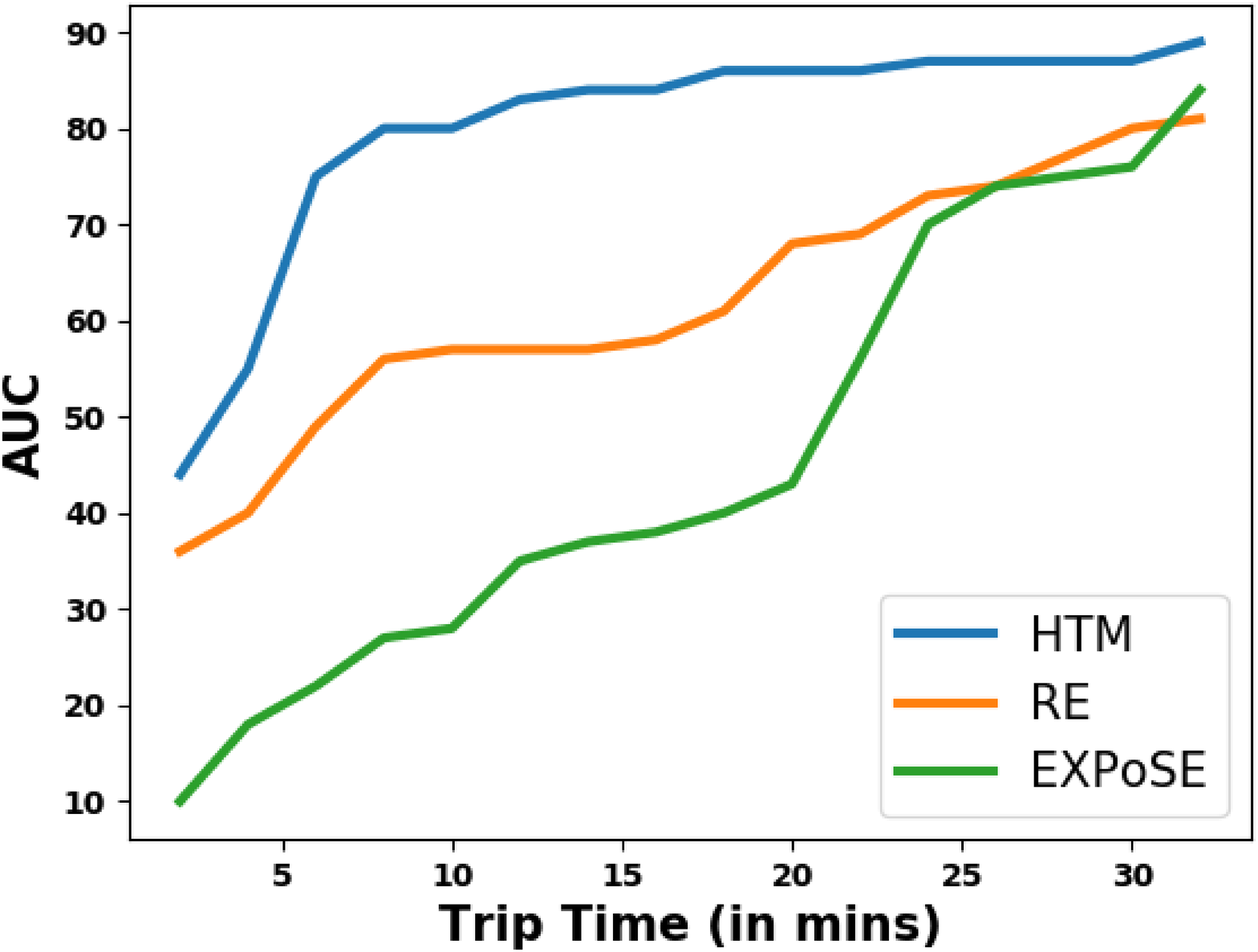}
  \caption{Change in AUC with percentage of data used for training for all the competing models.}
  \label{htmcompete}
\end{minipage}
\end{figure}

\subsubsection{Impact of data stream size}
The prime reason behind using HTM, as discussed before, was its ability to work well with sparse data, which could be utilized to detect anomalies early on the trip. In order to check how early can the system catch such anomalies, we performed another experiment. We trained the model with the data available from only $2 mins$ of the trip up to $16 mins$ and then tested over the incoming stream of the data for trips more than $20 mins$ (such that we can test for at least $20\%$ of the total data). We measure \textit{Area under the Receiver Operating Characteristics (ROC) Curve} (AUC), which indicates how much the model is capable of separating anomalies in the given input data stream. We use AUC to measure the performance of the HTM model as the number of anomalous cases are much less compared to the non-anomalous cases. The AUC results averaged over all the trips for the three features (jerkiness, speed, and congestion) are given in Fig.~\ref{htmmin}. As is evident, we get good AUC score even at $8 mins$ of trip time, which almost stabilizes after $12 mins$. Moreover, it reaches above $80\%$ for speed and jerkiness at $10 mins$. Thus, for all our experiments, we start predicting after the first $10 mins$ of the trip gets completed.

Fig.~\ref{htmcompete} shows the AUC values averaged over all the three features for the three models. Compared to the competing models, HTM based model provides better accuracy quite early on a trip. Relative Entropy also provides acceptable accuracy over the trip. However, EXPoSE improves only when it receives considerable data for online training and eventually is almost equal to Relative Entropy and comparable to HTM. The figure indicates that HTM converges much faster compared to Relative Entropy and EXPoSe, and therefore much more suitable in a real-time prediction problem. 

\begin{table}[!ht]
\begin{minipage}{0.48\linewidth}
\centering
\caption{Change in AUC on varying $W$ and $W'$. Here we have only shown the best 5 combinations.}
\label{ww}
\begin{tabular}{|l|l|l|l|l|}
\hline
\textbf{W}    & \textbf{W'} & \textbf{Speed} & \textbf{Jerkiness} & \textbf{Congestion} \\ \hline
3000          & 5           & 0.81           & 0.73               & 0.71                \\ \hline
\textbf{4000} & \textbf{10} & \textbf{0.83}  & \textbf{0.85}      & \textbf{0.76}       \\ \hline
4000          & 15          & 0.57           & 0.56               & 0.8                 \\ \hline
5000          & 5           & 0.61           & 0.84               & 0.53                \\ \hline
6000          & 10          & 0.78           & 0.51               & 0.66                \\ \hline
\end{tabular}
\end{minipage}
\hfill 
\begin{minipage}{0.48\linewidth}
\centering
\caption{Comparison of AUC results with existing models. HTM performs better than RE and EXPoSE.}
\label{htmauc}
\begin{tabular}{|c|c|c|c|}
\hline
\textbf{Model}  & \textbf{Speed} & \textbf{Jerkiness} & \textbf{Congestion} \\ \hline
\textbf{RE}     & 0.65           & 0.74               & 0.54                \\ \hline
\textbf{EXPoSE} & 0.4           & 0.47                & 0.21                \\ \hline
\textbf{HTM}    & 0.83           & 0.86               & 0.78                \\ \hline
\end{tabular}
\end{minipage}
\end{table}

\subsubsection{Anomaly detection performance}
Table~\ref{htmauc} gives the results of the mean AUC for anomaly detection module for all three features compared to the existing models. The online training is done for the first $10 minutes$ of the trip, after which the simultaneous learning and prediction phase starts. EXPoSE, being highly dependent on the size of the dataset, provides inferior results as the data it receives on the first $10$ minutes of the trip is not sufficient for its convergence, as we have seen earlier. The entropy-based approach also performs poorly as it is known to provide comparatively poor results when the features show both spatial as well as temporal variation at the same time.


\subsection{Performance Evaluation: Comfort Level Predictor}
The model was trained using the data collected during the data collection phase (Section~\ref{datacollection}). The Feature Extractor calculated all the features required by the model. The discomfort likelihoods were then obtained from the HTM model. Following this, discomfort likelihood scores for the three spatio-temporal features (\textit{speed, congestion, and jerkiness}) along with the three instantaneous features ($T_t, d_t$, and $Z$), were fed in the model along with the labels obtained from the commuters. The model was then trained using a loss function for softmax regression~\cite{heckerman1997models} with $60\%$ data for training and $20\%$ for validation. The remaining $20\%$ data was used for testing.

\begin{figure}[!ht]
  \centering
  \includegraphics[width=0.7\linewidth]{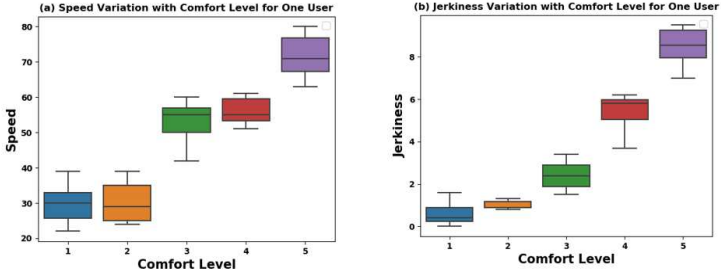}
  \caption{Variation of a feature with respect to comfort level of a user who is primarily affected by the same. (1 - Most Comfortable, 5 - Most Uncomfortable)}
  \label{ftrimpact}
\end{figure}

We evaluate the trained MTL-NN model over the data collected from the ten volunteers who took part in the experiments in Section~\ref{datacollection}. The discomfort likelihood scores are obtained from the HTM module for all the three spatio-temporal features, and the remaining three instantaneous features are directly obtained from the Feature Extractor. As we have discussed earlier, \our labels the data points on a scale from $1$ (\textit{highly comfortable}) to $5$ (\textit{highly uncomfortable}); however, as most of the data points are labeled between 1-3, considering this unbalanced dataset, we compute the AUC\revision{~\cite{huang2005using,galar2011review}}. Moreover, in light of the multi-class classification, we utilize a forced binary classification using the \textit{one-versus-all} approach. For instance, we consider $1$ as the success class and all other combined as the failure class. We then plot the ROC for all these separate instances and give the AUC result aggregated over the number of classes. The results of the aggregated AUC and for the five instances are given in Table~\ref{mtlauc}, where we obtained an average AUC score of $0.876$. It can be observed that the AUC for label 5 (\textit{highly uncomfortable}) is the highest, which can mostly be linked to extreme scenarios that cause high discomfort for a commuter at a personal level and would have quite distinctive characteristics compared to other labels. In Fig~\ref{ftrimpact}, we plot the variation for speed and jerkiness with respect to comfort level for two users. One of them is highly impacted by the speed variations while the other due to jerkiness. As can be seen, the characteristic for level 5 is quite extreme and easily distinguishable from the other labels in both the scenarios.

\begin{table}[!ht]
\scriptsize
\begin{minipage}{0.48\linewidth}
\centering
\caption{AUC Scores for all labels}
\label{mtlauc}
\begin{tabular}{c|c|c|c|c|c|c|}
\cline{2-7}
                                   & \textbf{1} & \textbf{2} & \textbf{3} & \textbf{4} & \textbf{5} & \textbf{Average} \\ \hline
\multicolumn{1}{|c|}{\textbf{AUC}} & 0.876      & 0.864      & 0.861      & 0.884      & 0.893      & 0.876            \\ \hline
\end{tabular}
\end{minipage}
\hfill
\begin{minipage}{0.48\linewidth}
\centering
\caption{Sobol Indices (TOI) for the Six Features.}
\label{sobol}
\scriptsize
\begin{tabular}{|l|l|l|l|l|l|l|}
\hline
\textbf{Feature}                & \textbf{$\mathcal{L}_t^c$} & \textbf{$T$} & \textbf{$\mathcal{L}_t^s$} & \textbf{$\mathcal{L}_t^j$} & \textbf{$t$} & \textbf{$d$} \\ \hline
\textbf{TOI}    & 0.89       & 0.85       & 0.80       & 0.75       & 0.71      & 0.66      \\ \hline
\end{tabular}
\end{minipage}
\end{table}

\begin{figure}[!ht]
\begin{minipage}{0.48\linewidth}
  \centering
  \includegraphics[width=\linewidth]{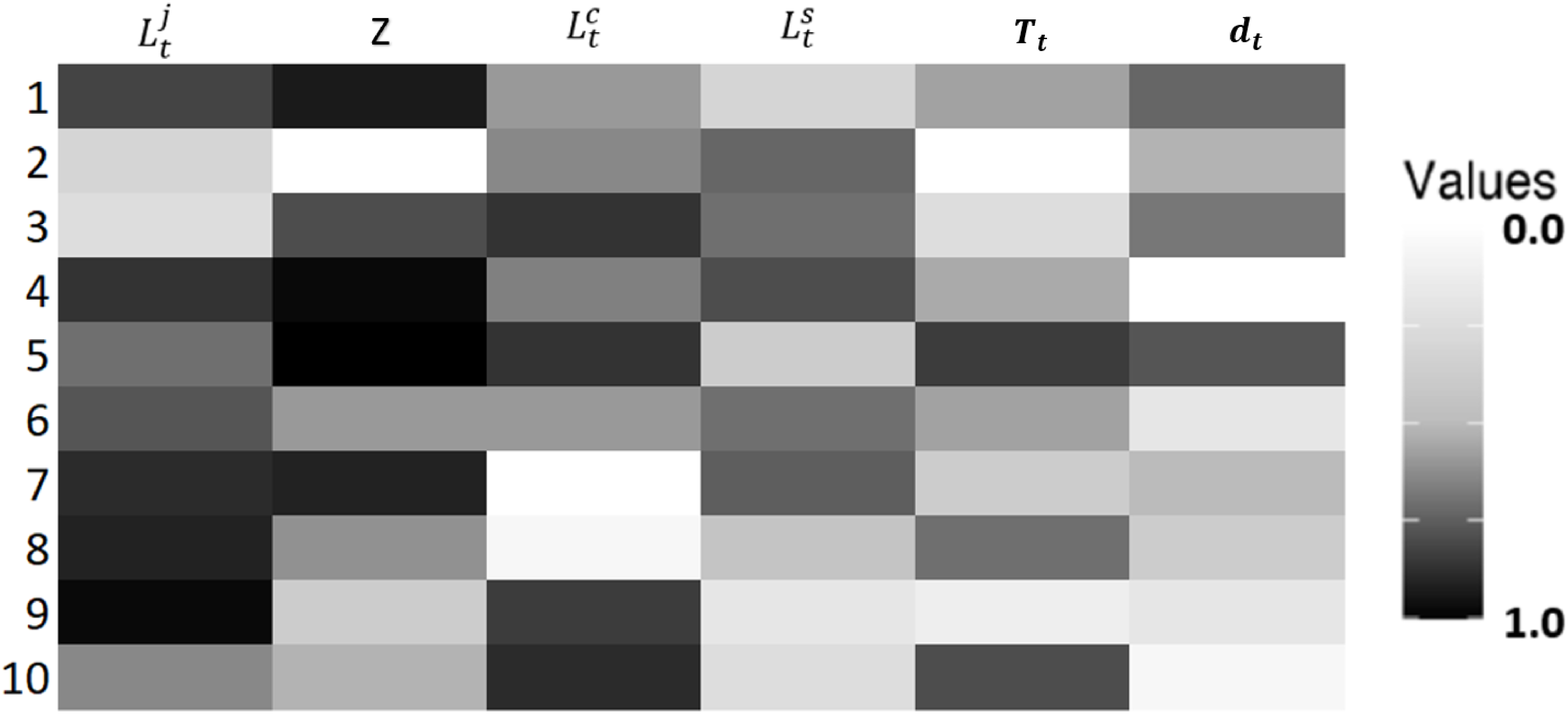}
  \caption{Impact of features on 10 users. Darker value implies higher impact of the feature on a user.}
  \label{heat}
\end{minipage}
\hfill 
\begin{minipage}{0.48\linewidth}
  \centering
  \includegraphics[width=\linewidth]{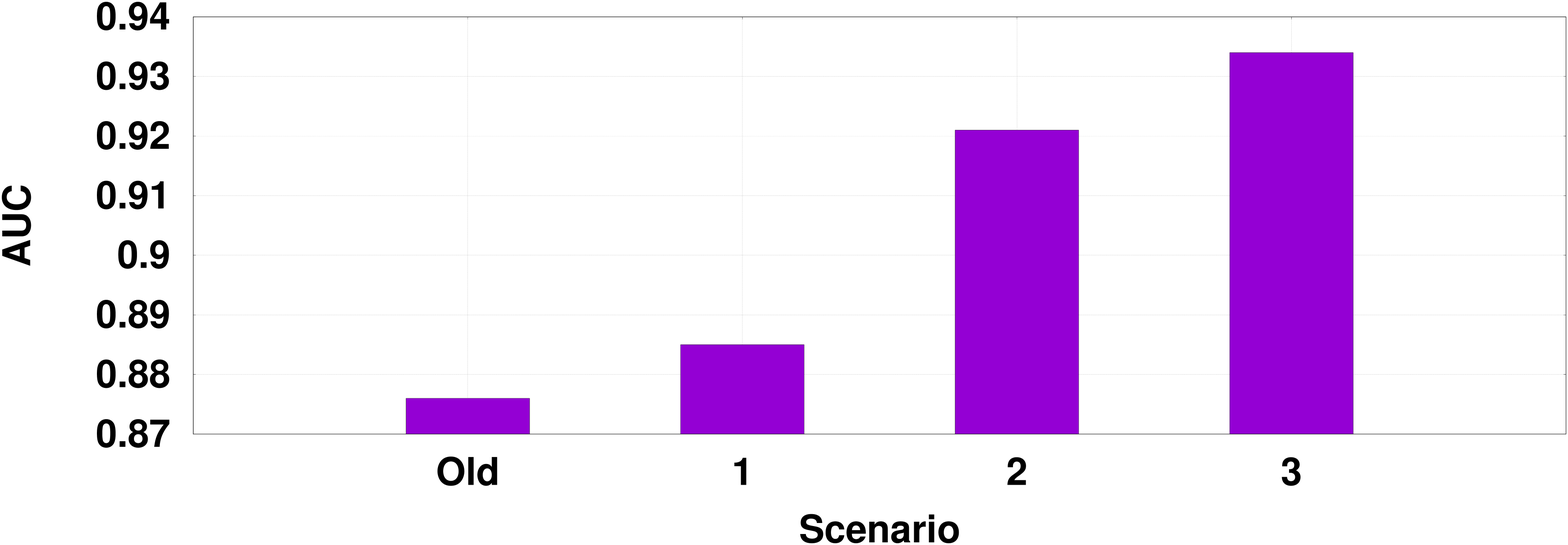}
  \caption{Impact of data augmentation. Scenarios: (1) Data from any one new commuter (2) Data from all new commuters (3) New data from all commuters}
  \label{dataaug}
\end{minipage}
\end{figure}

In order to obtain the classification importance of each of the six features, we performed sensitivity analysis~\cite{saltelli2000sensitivity}. Sensitivity analysis is the study of relative interaction of different input factors on the model output. We used \textit{Sobol Total Order Indices}~\cite{sobol1993sensitivity} to perform sensitivity analysis as it converges to the exact relative contributions and interactions of the input factors with respect to the variability in the output. The results are given in Table~\ref{sobol}, and we observe that the total order confidence is below $10\%$ for each feature, thus confirming that the sample size provided is sufficient for the analysis and the measured indices are significant. We observe that \textit{congestion} followed by the \textit{time of the day} has the highest impact on the discomfort a commuter feels, which also seems intuitive. Congestion is associated with long waiting times, and taking a trip at night or early morning usually would make a commuter more uncomfortable with even a small variations in the driving. However, this need not be true for all, as is evident from Fig.~\ref{heat} where we have shown the impact of different features on 10 randomly chosen users. This brings out the personalization aspect clearly as we can see that each commuter is affected by different features differently.

\subsubsection{Impact of data augmentation}
We also tried to observe the impact of data augmentation on the model, where we retrained the model over the data collected during the testing phase of the experiments. Here, the commuter was asked to provide their labels whenever the model was nearly ambiguous (decided based on comparable classification probability between two or more classes, as discussed earlier) about predicting the label. In order to perform this experiment, we observed the results when we received new data in three scenarios;
(1) \textbf{One new app user:} As mentioned earlier, only five of the $30$ users were involved in the data collection experiment described in Section~\ref{datacollection}. Whenever one of the new users were polled for feedback, the data was tagged as to be obtained from a new user. In this scenario, we only considered the impact on the model when we added data from only one of these $25$ new commuters to the existing dataset. 
(2) \textbf{All new app users:} In this scenario, data from any new user was added to the dataset, and the corresponding impact on the model was observed.
(3) \textbf{All app users:} In this experiment, we collected feedback from all the $30$ users who had used the application.

Following this, we trained and tested the model with a $60-20-20$ split for training, validation, and testing. It should be noted that for the first scenario, the test was done for all new app users, and the final AUC was calculated as an average over the result of adding data of any one new user to the existing data. As is observed from Fig~\ref{dataaug}, data from one new user improves the results but not much. However, adding data from all the $25$ new users considerably improves the model. Nevertheless, once the model has learned over all 30 users, adding new data over this, though improves the performance, but not considerably.

\revision{We also noted the instances when the model requested for user feedback owing to nearly ambiguous prediction. There were two scenarios when such drop in confidence could occur;}

\revision{\noindent{\textbf{Existing Commuter:}} In this scenario, we consider any existing user, who has used the app for at least a week, for conducting the experiment and observe that on average only for 5\% of cases, there is a drop in confidence. This is mostly attributed due to the (rare) changes in commuter preference in a trip, for almost similar conditions (say, in a trip, she initially preferred moderate speed, however, at the last leg of the trip, preferred high speed to quickly reach the destination).}

\revision{\noindent{\textbf{New Commuter:}} In this scenario, we consider the case where a new user joined the experiment. This new commuter’s comfort labels are initially detected from the existing model, trained on few existing commuters, exhibiting similarity with the new commuter (similarity is handled by the MTL-NN). Evidently, the proposed model makes mistakes for those new commuters, exhibiting drop in confidence in indicator vectors. Precisely, frequent retraining was required initially for these new commuters (average of ~35\% labels requiring a commuter feedback), but it gradually decreased with time.}

\subsection{Application: Driver Rating System based on \our}
In this subsection, we provide a prototype application where we use \our to assist the commuters. It should be noted, though, that the following application is just a \textit{proof-of-concept} to show the utility of \our and can be further modified and looked into as a separate research problem.

Ratings are an essential aspect of companies like Uber/Lyft, which affect the driver's commission, number of rides, and in pressing cases losing their job~\cite{uberdeactivate}. However, commuters are usually conflicted when giving a rating to a driver unless the ride has been poor~\cite{uberrate, uberrate2}. This, in turn, profoundly affects the driver as well as the company reputation. An application that takes cues from the driving and provides a suitable rating to the driver would thus be quite useful. \our could be used as an excellent framework for such an application. We added a module to our application (Fig~\ref{app}(b)), which performs an averaging over the comfort rating throughout the trip to rate the driver. It also shows the impact of speed, congestion, and jerkiness over the complete trip comfort averaged over individual values. As can be observed in the figure, we also asked the commuters to provide a comfort rating to the ride, which was stored in our server as ground truth value.

In order to calculate the agreement between the calculated and user ratings, we use Kendall's coefficient of concordance ($\mathbb{W}$)~\cite{kendall1939problem} which is a good metric for such 5-point rating scales~\cite{taureason}. This is calculated as $\mathbb{W} = \frac{12 \sum_{i=1}^{n} (R_i - \overline{R})^2}{m^2(n^3 - n)}$, where $R_i$ is the total rank given to rating $i$, $\overline{R}$ is the mean of $R_i$ while $m$ and $n$ are the numbers of competing rating systems and number of ratings respectively. Here $m = 2$ and $n = 5$. Calculating over all the responses, we observed a $\mathbb{W}$ value of $0.79$, which is considered as a good agreement as per the existing literature~\cite{Wscore}.

\section{Discussion}\label{discussion}
Although \our shows considerable promise as a system to assess commuter comfort at a personalized level which could be utilized by many other services, in this section we discuss some limitations and future directions to improve the overall system.

\subsection{Incorporating Additional Features for Model Improvement}
\our  focuses on two generic feature classes -- (i) instantaneous (time of the day, distance traveled, and time traveled) and (ii) spatio-temporal features (speed, congestion, and jerkiness), rather than specific feature variations, and use directly available quantitative features to develop the model. However there could be several non-quantifiable features which do impact the commuter's comfort. For instance, the personalized features like if the commuter is in a hurry, weather condition, laptop or phone usage, etc. are more qualitative in nature. A possible direction could be to take such information as an input from the user. For instance, expected travel time could be an input from the user which could be normalized based on the average travel time on the route to measure urgency; the OpenWeatherMap Weather API could be used to get the weather condition on a three-point-scale~\cite{vermapercom}; binary inputs could be taken from the user for usage of laptop or phone. The MTL model that we have used for comfort level prediction, is generic and is expected to provide proper predictions when including such features with suitable quantitative mapping.


\subsection{Improving the Rating System}
The rating system discussed in this paper is a simple \textit{proof-of-concept} to show the utility of \our. The main goal of this work is to develop a methodology to connect commuter’s comfort with the driver rating system while computing the comfort solely from the travel parameters without explicitly asking the commuter and thus eliminating a rating bias. However, any rating system has a primary linkage with the business policy of the cab companies, therefore, the cab companies can use a more sophisticated rating system, which might even vary across cab companies, while considering the commuter’s comfort as one of the important parameters. 

\subsection{Other Applications of \our}
There are several other directions we could look into as potential applications of \our. For the commuter, an application could provide more information from historical ratings or route based comfort information. Another application could make available to the driver the commuter’s comfort state and also the reason behind that. That way the driver could take necessary steps if possible to improve the commuter's comfort. Moreover, if both are using the application, a profile sharing based on the application could be done such that the driver could take better driving decisions as well as the commuter is ready for the ride. This could be further extended to include a driver recommendation system based on profile matching. 
\section{Conclusion} \label{conclusion}
In recent times, there has been an increasing demand for comfortable ride-sharing options like Uber, Lyft, etc. in contrast to public transport. As these ride-sharing companies hugely rely on the ratings the drivers received from the commuters, it has become imminent to maintain the comfort level for a commuter taking the ride. In light of this, we develop a system \our, which understands the comfort needs of a commuter at a personalized level and computes whether a specific driving style at a time on the trip is causing discomfort to the commuter. Based on an online survey and pilot study, we understand what features could affect the comfort of a commuter. We then use a Hierarchical Temporal Memory and Multi-task learning-based model to compute if any change in the distribution of three spatial time-series features-- speed, jerkiness, and congestion-- along with other static trip information is causing discomfort to a commuter and to what level.

Furthermore, we also add another feature in \our, which checks if the current computation of comfort level is near ambiguous and requests the commuter for feedback, which improves the dataset on which further training could make the model robust and scalable to new and existing users both. Thorough experiments with \our shows that it not only computes the comfort levels effectively but could also understand at what level does a feature affects a particular commuter's comfort. Thus, efficiently capturing the personal comfort needs of the commuter. Such a system, which computes commuter discomfort at a personalized level, could be utilized for several applications like driver rating, alerting a driver of a commuter's discomfort, assigning drivers to commuter based on her comfort profile, etc. We have built a comfort rating application to show the utility of the comfort calculation framework. Further detailed research in this line could help build much more efficient and similar applications utilizing the perception of commuter comfort during a cab-ride.

\bibliographystyle{ACM-Reference-Format}
\bibliography{cscw}

\end{document}